\documentclass[iop,numberedappendix]{emulateapj}
\usepackage{amsmath}
\usepackage{bm,xspace}
\usepackage{times}
\newcommand{\Eq}[1]{Equation~(\ref{#1})}
\newcommand{\EQ}{\begin{equation}}
\newcommand{\EN}{\end{equation}}

\newcommand{\St}{\text{St}}

\newcommand{\SR}{SR\xspace}
\newcommand{\pice}{p_{\text{ice}}}
\newcommand{\Torb}{\triangle T_{\text{Orb}}}
\newcommand{\Lorb}{\triangle L_{\text{Orb}}}
\newcommand{\Rorb}{\triangle R_{\text{Orb}}}
\graphicspath{{.}{./Figures/}}

\date{\today,~ $ $Revision: 1.20 $ $}
\begin{document}

\title{FU Orionis outbursts, preferential recondensation of water ice, and the formation of giant planets}
\shorttitle{Preferential recondensation of water ice}

\author{Alexander Hubbard\altaffilmark{1}}
\altaffiltext{1}{American Museum of Natural History, New York, NY, USA}
\email{{\tt ahubbard@amnh.org}}

\begin{abstract}
Ices, including water ice, prefer to recondense onto pre-existing nuclei rather than spontaneously forming grains from
a cloud of vapor. Interestingly, different potential recondensation nuclei have very different propensities to actually nucleate water ice
at the temperatures associated with freeze-out in protoplanetary discs. Therefore, if a region in a disc is warmed and then recooled,
water vapor should not be expected to refreeze evenly onto all available grains. Instead it will preferentially recondense onto the most
favorable grains. When the recooling is slow enough, only the most favorable grains will nucleate ice, allowing them to recondense thick
ice mantles. We quantify the conditions for preferential recondensation to rapidly create pebble-sized grains in protoplanetary discs and show
that FU Orionis type outbursts have the appropriate cooling rates to drive pebble creation in a band about 5 astronomical units wide
outside of the quiescent frost line from approximately Jupiter's orbit to Saturn's (about $4$ to $10$\,au).
Those pebbles could be of the appropriate size to proceed to planetesimal formation via the Streaming Instability,
or to contribute to the growth of planetesimals through pebble accretion. We suggest that this phenomenon contributed to the formation
of the gas giants in our own Solar System.
\end{abstract}


%%%%%%%%%%%%%%%%%%%%%%%

\section{Introduction}
Ices are a major player in planet formation.
In decreasing order of condensation temperature,
rocky material, water ice, and all other ices each make up about $0.5\%$ of a protoplanetary disc's total mass, and one
third of the potential solid mass \citep{2003ApJ...591.1220L}.
Accordingly, just outside the water frost line where water is in solid form, the disc has about twice the material available
to participate in the growth of solids as inside the frost line.
Ice-rimmed dust is also stickier and more collision resilient than
bare silicates \citep{1997ApJ...480..647D}, further promoting the collisional growth of solids.
The temperature at which the water ice saturation vapor pressure equals the water vapor partial pressure depends
on the local number density of water molecules.

Nonetheless, cooling from $T=170$\,K to $T=166$\,K more than halves the saturation vapor pressure \citep{2005QJRMS.131.1539M}.
Thus it is reasonable to approximate that the temperature window in which water is not
effectively entirely in solid or gaseous phases is extremely narrow,
justifying the traditional assumption that the window occurs at $T=170$\,K  \citep{2000ApJ...528..995S} although
this is complicated by disc vertical structure \citep{2004M&PS...39.1859P}.
Disc regions at larger orbital separations have less ambient material, and therefore
must be colder to condense water ice, but the difference in condensation temperatures is small enough that
the radial temperature gradient dominates, and the fraction of water condensed rapidly approaches unity outside of the frost line.

We can assume that water vapor is in equilibrium with ice-mantled dust grains
as long as the system evolves slowly enough and, crucially, as long as a significant portion of the water ice remains condensed and exposed.
If a parcel of gas and ice-mantled dust experiences a temperature fluctuation
sufficient to evaporate the ice before cooling, equilibrium cannot be assumed.
Homogenous freezing, spontaneous freezing in the absence of preexisting nuclei, is more difficult
than inhomogenous freezing onto existing potential ice nuclei (henceforth IN), and does not
occur under the same temperature and pressure conditions as evaporation \citep{2000Natur.406..611K}.
Interestingly however, not all IN are created equal \citep{Cziczo1320}:
at temperatures associated with water freezing in protoplanetary discs, i.e.~$T<170$\,K, IN of differing qualities
can require saturation ratios \SR of factors of several to begin nucleating ice \citep{2013JGRE..118.1945C}.
Ice is the best surface at condensing more ice, so once mantles are accreted, the differences between ice mantled INs vanish.

If, then, a parcel of gas and ice-mantled dust is heated sufficiently to evaporate all the ice, and then recooled slowly, we can expect
the most favorable potential INs to accrete ice mantles first. With those mantles in place, the favored few grains will maintain equilibrium
between their water ice surfaces and the water vapor, i.e.~a saturation ratio \SR$=1$, preventing water ice from recondensing on the other
grains even if they had originally possessed ice mantles. We refer to this process as \emph{preferential recondensation}.
 If, on the other hand, the parcel is cooled sufficiently rapidly, the rise in \SR due to cooling
would outpace the drop in \SR due to condensation enough for the next tier of INs to also nucleate ice.

One scenario where we would expect preferential recondensation to occur with important consequences
is the aftermath of a major accretion event such
as an FU Orionis type outburst \citep{1996ARA&A..34..207H}. FU Orionis events occur early in a protostar's existence
and lead to significant disc heating \citep{2016arXiv160703757C}. 
As we will show, during an FU Orionis outburst the entire radial belt from
about $4$\,au to about $10$\,au could host significant preferential recondensation.
By restricting the recondensation to a small subset of dust grains, those grains would grow to sizes associated with
both planetesimal formation and pebble accretion \citep{2007Natur.448.1022J,2012A&A...544A..32L,2015A&A...579A..43C}.
This provides a pathway to rapidly triggering and promoting the formation
of giant planets in the outer disc where densities are low and dust coagulation is unlikely to proceed apace.

\section{Model}

As might be expected, there is a vast literature discussing ice formation in the context of terrestrial cloud formation, far beyond
the author's expertise and impossible to summarize \citep{cantrell2005production,hoose2012heterogeneous}. Protoplanetary discs are however expected to be in a different
regime than our atmosphere, with nearly unity IN to water density ratios, and correspondingly large
effective IN number densities. Note that, excepting explicit densities of solid grains, any densities we refer to are densities per unit volume
of protoplanetary disc gas plus solids. Protoplanetary discs also have sufficiently slow evolution time scales, sufficient turbulent
mixing, and sufficiently long molecular mean-free-paths
that freeze-out is not expected to be diffusion limited: turbulent mixing and molecular diffusion replenish
the water vapor near a dust grain as fast as it is lost to condensation.
Further, grains cannot rapidly move out of condensation regions
without first growing to meaningful size.

Those large potential IN to water density ratios, combined with relatively slow temperature fluctuations,
make homogenous freezing a negligible phenomenon in protoplanetary discs.
In the case of the terrestrial atmosphere, one is often in the situation where multiple condensation conditions are met,
and which path is taken, such as continued homogenous condensation, or inhomogenous condensation onto newly homogenously
formed INs is a kinetic question \citep{JGRD:JGRD15209}. We can instead assume that the condensation saturation ratios \SR are significantly
separated for different INs, and that those \SR are in turn all significantly below the \SR required for homogenous freezing.

In this section we derive a basic model for the kinetics which uses those simplifying assumptions. 
By assuming that different IN have significantly different \SR critical values to begin nucleating ice, we allow condensation to be restricted to
a sub-set of IN even if the \SR value temporarily rises. 
Thus, we only need to find the conditions required to avoid large \SR fluctuations.
More sophisticated models take the time derivation of the saturation ratio \SR, e.g.~\cite{JGRD:JGRD8827}.
Making effective use of those models would however require a possessing a detailed model of the distribution of the critical \SR values for the
potential IN actually present.
 
 \subsection{Comparing time scales}
 
The saturation pressure of water vapor over ice is approximately \citep{2005QJRMS.131.1539M}:
\EQ
\pice \simeq \exp(28.9074-6143.7/T), \label{pice}
\EN
where $\pice$ and $T$ are measured in Pa and K respectively.
We can use \Eq{pice} to define
\EQ
\tau_p(\partial_t T)= \frac{\pice}{\partial_t \pice} = \frac{\pice}{\partial_T \pice} \frac{1}{\partial_t T} \simeq \frac{T^2}{6143.7} 
 \left(\partial_t T\right)^{-1}, \label{taup}
\EN
the time scale for the vapor pressure to change as a function of the heating or cooling rates.

We can compare this
time scale $\tau_p$ to an equilibration time scale between ice and vapor, which we will
quantify through the time required for a vapor water molecule to encounter and expect to stick to an icy target:
\EQ
\tau_c = \frac{1}{\alpha \sigma n_s v_w}, \label{tauC1}
\EN
where $\alpha$ is the accommodation coefficient, $\sigma=\pi a^2$ is the collisional cross-section of the icy grains assumed to be spheres of radius $a$,
$n_s$ is the number density of icy grains and
\EQ
v_w = \sqrt{\frac{k_B T}{m_w}}
\EN
is the thermal speed of a water molecule.
The appropriate accommodation coefficient is unclear. Modeling observed cloud formation suggests low values potentially
below $10^{-2}$, but recent laboratory studies have found $\alpha \gtrsim 0.5$, which value we will use \citep{acp-13-4451-2013}.

\Eq{tauC1} estimates the time-scale on which water molecules freeze out onto ice grains, which means
that $\tau_c$ is also the time-scale on which the ice partial
pressure drops due to recondensation (any drop in partial pressure due to cooling is negligible for our purposes).
In a cooling disc,  recondensation is rapid enough to keep water ice in rough equilibrium
with icy surfaces as long as $\tau_c < \tau_p$. If on the other hand $\tau_c > \tau_p$, then recondensation will lag and the
\SR will rise, triggering recondensation onto less and less favorable dust grains, causing $\tau_c$ to drop over time (more grains to
recondense on). Once exactly enough dust grains begin condensing water that $\tau_c=\tau_p$, equilibrium can be maintained,
and new grains will not join in.

Assuming ice-vapor equilibration ($\tau_c \lesssim \tau_p$), it takes very little cooling
for nearly complete freezing, allowing us to approximate
\EQ
n_s \times \frac{4\pi}{3} \rho_s a^3 = \rho_w, \label{ns}
\EN
where $\rho_s$ is the approximate solid density of our ice-mantled grains
and $\rho_w$ is the fluid density of water molecules in the disc. Note that we have assumed that
all the ice nucleating grains are of the same size, and have nucleated sufficient ice to dominate their mass and radius.

Combining Equations~(\ref{tauC1}) and (\ref{ns}), we arrive at
\EQ
\tau_c = \frac{4 \rho_s a}{3 \rho_w v_w} \alpha^{-1}. \label{tauC2}
\EN
Denoting the water-to-gas mass ratio as $\epsilon= \rho_w/\rho_g$, we can rewrite \Eq{tauC2} as
\EQ
\tau_c = \frac {4}{3\epsilon} \frac{v_{th}}{v_w} \sqrt{\frac{8}{\pi}}\frac{\tau_E}{\alpha}, \label{tauC3}
\EN
where
\EQ
\tau_E = \sqrt{\frac{\pi}{8}} \frac{a \rho_s}{\rho_g v_{th}} \label{tauE}
\EN
is the Epstein regime drag time scale of the dust.
The thermal speed of the gas is
\EQ
v_{th}=  \sqrt{\frac{k_B T}{m_g}} = \sqrt{\frac{m_w}{m_g}} v_w  \simeq 3 v_w
\EN
where $m_g \sim 2$\,amu is the gas mean molecular mass.

We can use Equations~(\ref{taup}) and (\ref{tauC3}) to write the condition $\tau_c = \tau_p$ as
\EQ
\Torb St \simeq \alpha \frac{T^2}{6143.7} \frac{3 \pi \epsilon}{2} \sqrt{\frac{\pi}{8}} \frac{v_w}{v_{th}} \simeq 0.01,
\label{modelmodel}
\EN
where we have used $\epsilon \simeq 0.005$, $v_{th} \simeq 3 v_w$, and $T = 160$\,K; and $\Torb$ is the change in
temperature in degrees Kelvin over a local orbital period. \Eq{modelmodel}
estimates the largest Stokes number $St$ at which icy grains can recondense ice fast enough to maintain equilibrium
with water vapor
for a cooling rate defined through $\Torb$. Alternatively, it defines the fastest cooling rate $\Torb$ at which icy
grains with Stokes number $\St$ can maintain equilibrium with water vapor.

In \Eq{modelmodel} the drag time has been non-dimensionalized
with the local orbital frequency $\Omega$ through the Stokes number of the dust:
\EQ
St\equiv\tau_E \Omega.
\EN
If a region in the disc heats enough to evaporate the ice and then cools at a rate
of about $1$\,K per orbit, the water vapor is expected to recondense onto a small number of INs,
forming dust grains with $St \sim 0.01$, large enough to have significant consequences for planet formation
\citep{2007Natur.448.1022J,2012A&A...544A..32L,2015A&A...579A..43C}.

\subsection{Latent heat}

There is a further complication to estimating $\Torb$: the latent heat released
by water freezing is significant. From \cite{2005QJRMS.131.1539M}, we have
\EQ
L_w \simeq 2.7 \times 10^{10} \text{ erg g}^{-1}.
\EN
Writing
\EQ
\frac{k_B \triangle T}{m_g} = \epsilon L_w,  
\EN
we find that the latent heat is sufficient to correspond to a temperature change of
\EQ
\triangle T \simeq 3 \text{ K}.
\EN
At a background temperature of $T\sim 160$\,K \Eq{pice} implies that
the $3$\,K temperature difference provided by condensing water vapor in a protoplanetary disc is sufficient to
to halve the saturation water vapor pressure. If cooling is sufficiently slow for \Eq{modelmodel} to have
significant implications, latent heat could meaningfully further slow the cooling rate.

\section{Cooling regimes}

The details of ice deposition only matter when condensation occurs but is not total, i.e.~near a water ice frost line.
We examine two cases, static frost lines, and evolving frost lines.

\subsection{Static frost line}

In the case of a static frost line, we have $\Torb=0$, which in conjunction with \Eq{modelmodel} would seem
to imply the growth of very large ice rimmed grains indeed. However, water vapor will only recondense
after being transported across the frost line. Turbulence mixes the water vapor into regions with pre-existing water
ice rimmed grains, moving radially at most a turbulent length scale within a turbulent time scale assumed to be approximately
the orbital time scale \citep{2006A&A...452..751F}.

Turbulence has a length scale
\EQ
l_t \simeq \sqrt{\alpha_{SS}} H
\EN
where $\alpha_{SS}$ is the Shakura-Sunyaev $\alpha_{SS}$ \citep{1973A&A....24..337S}, and $H$ the local scale height. We expect 
the disc background temperature
to scale as $R^{-1/2}$ as in a \cite{1981PThPS..70...35H} minimum mass solar nebula (MMSN), so moving one turbulent length scale would correspond to about
\EQ
\frac{\delta T}{T} \simeq \frac{\sqrt{\alpha_{SS}}}{2}\frac{H}{R} \simeq 8 \times 10^{-4} \label{deltaT}
\EN
where we have used $\alpha_{SS} \sim 10^{-3}$ and $H/R \sim 0.05$. At a background temperature of $T=160$\,K
\Eq{deltaT} implies
\EQ
\delta T \sim 0.13\,\text{K}.
\EN
Thus it is unlikely that water vapor would be able to freeze out, even in an inhomogenous manner, except
onto pre-existing ice-mantled dust grains. Even in that case, the approximation that freeze-out or evaporation is total does not apply
for such a small $\delta T$: while narrow, frost lines are clearly broader than turbulent length scales.

Turbulently mixing two equal volumes of protoplanetary gas and dust
outside a frost line, one with ice condensed, and the other, evaporated, will result in the water vapor only
recondensing on the pre-existing ice-mantled grains. Even complete recondensation would 
at most double the mass of the icy grains.
From \Eq{tauE} we can see that this would only increase the icy grain stopping time by a factor between $2^{1/3}$ (solid grains
growing at constant density)
and $2$ (highly porous grains growing at constant radius).

\cite{2013A&A...552A.137R} showed however that a small number of icy particles outside a frost line will remain there
long enough to be mixed into water vapor rich parcels of gas turbulently transported outside the frost line several times,
allowing them to grow to significant size. Considerations of differing
condensation nuclei qualities only strengthens this conclusion by arguing that the bare IN also carried in the water vapor rich parcels
are unlikely to begin to recondense ice before the icy grains freshly mixed into the parcels can do so.

\subsection{Evolving frost line}

\subsubsection{Cooling discs}

A more interesting case from our perspective is a cooling disc whose frost line is contracting, causing
large radial regions to experience freeze-out. As long as the frost line retreats sufficiently, it will
eventually reveal completely dry grains. When global scale cooling rapidly enough \Eq{ns} applies, and hence \Eq{modelmodel} holds.
That will certainly occur when the frost line retreats faster than turbulence can mix material from both sides of the line.
Taking advantage of the very strong temperature dependence of the vapor pressure,
we quantify that limit by requiring the frost line to retreat more than a turbulent length scale over a turbulent period, which we
estimate as an orbit \citep{2006A&A...452..751F}.

To accurately determine the radial
temperature profile of a disc the full radiative transport equations must be taken into account, and discs possess
vertically varying thermal structures \citep{2002A&A...389..464D}. The strong dust dependence of radiative transport 
further complicates the issue in the case of preferential recondensation, when the size of the dust varies rapidly
\citep{2009MNRAS.393.1377I}. In the case of a cooling disc, we can even expect a dust wall just outside of the frost line,
with a jump in the opacity. Given our uncertainties, we adopt the simplifying approximation that $T \propto R^{-1/2}$.

Thus, assuming a disc locally experiencing external irradiation (from the protostar or in the case of an FU Orionis event
the innermost rapidly accreting disc), we can write
\EQ
T^4 = \frac{A L}{R^2} \label{T_prop}
\EN
for some constant $A$, at an orbital position $R$ with external luminosity $L$. From \Eq{pice} the frost line temperature varies only slowly
with the local gas density. Assuming a frost temperature of $T_f \sim 160$\,K, quasi-constant as a function of radius the
frost radius is
\EQ
R_f = \frac{\sqrt{AL}}{T_f^2},
\EN
and the speed of its retreat is
\EQ
v_f=-\partial_t R_f = -\frac{1}{2T_f^2} \sqrt{\frac{A}{L}} \partial_t L = -\frac{R_f}{2} \partial_t \ln L.
\EN
The distance $\Rorb$ retreated in an orbit is then simply
\EQ
\Rorb = \frac{2\pi}{\Omega} v_f= \frac{R_f}{2} \frac{\Lorb}{L},
\EN
where $\Lorb$ is the chance in luminosity in one local orbit.
Requiring $\Rorb>l_t$ we arrive at the constraint that cooling must be faster than
\EQ 
\frac{\Lorb}{L}> 2\sqrt{\alpha_{SS}} \frac{H}{R}. \label{LHR}
\EN
At constant $R$ we can use \Eq{T_prop} to write
\EQ
\frac{\Torb}{T} = \frac 14 \frac{\Lorb}{L}. \label{TL}
\EN
We can use \Eq{LHR} to further determine that cooling outpacing turbulent mixing requires a dimming rate of
\EQ
\frac{\Lorb}{L} \gtrsim 0.003.
\EN
The condition for cooling to outpace mixing is that the external (inner disc or protostar) luminosity drops more than $0.3\%$ per local orbit.

\subsubsection{Applications to preferential recondensation}

Combining Equations~(\ref{modelmodel}), (\ref{LHR}), and (\ref{TL})
we arrive at
\EQ
St < 2.5 \times 10^{-4} \frac{R}{\sqrt{\alpha_{SS}}H}, \label{St_lim}
\EN
where we used $T \simeq 160$\,K.
For reasonable estimates of $H/R=0.05$ and $\alpha=10^{-3}$, \Eq{St_lim} becomes
\EQ
St \lesssim 0.08, \label{St_upper_limit}
\EN
implying that in the slow cooling limit, preferential recondensation can create quite large icy grains indeed.

We can also combine Equations~(\ref{modelmodel}) and (\ref{TL}), estimating $T=160$\,K, to write
\EQ
St = 2.5 \times 10^{-4} \frac{L}{\Lorb}.\label{cool_speed}
\EN
We are interested in preferential recondensation if it generates large grains. Arbitrarily setting the lower limit for large at $St\ge 10^{-3}$,
\Eq{cool_speed} implies $\Lorb/L<0.25$. Even extreme dimming rates can result in the condensation of respectably
large grains. \Eq{St_lim} implies that discs that are cooling sufficiently fast that the frost line outpaces turbulent mixing, but
not utterly outclasses it, are expected to see inhomogenous freeze-out on a small enough fraction of the ambient
potential INs so as to condense into large grains.

The range of dimming rates for which we expect preferential condensation onto favored INs to
result in large (here $St>10^{-3}$) grains is therefore
\EQ
3 \times 10^{-3} < \frac{\Lorb}{L} < 0.25, \label{Lorblimits}
\EN
although variations in $\alpha_{SS}$ or $H/R$ would adjust these dimming rate bounds.
Modest differences
in the thermal profile ($T \propto R^{-1/2}$) will adjust, but not qualitatively alter, \Eq{Lorblimits}.
The rates in \Eq{Lorblimits} have potential astrophysical implications, but, especially at the lower end
will require long-term monitoring surveys to fully explore (one orbit at
$4$\,au taking $8$ years for a  $1$M$_\odot$ star). In particular, the bounds match dimming rates associated with FU Orionis, the name-sake for
 FU Orinis type objects \citep{1996ARA&A..34..207H}. FU Orionis objects undergo violent accretion events, increasing
 in luminosity by around $6$ magnitudes, before dimming on a time scale of about a century.
 
 Recent observations have found that FU Orionis' continuum dimmed by $12\%$ over $12$ years, although there is a yet uncertain
 difference in the dimming rate between shorter and longer wavelengths, similar to previous estimates for BBW 76 and slower
 than the dimming of V1057 Cyg by a factor of about two
 \citep{2005MNRAS.361..942C,2006ApJ...648.1099G,JOEL16}. Increasing the luminosity of a Hayashi MMSN
 by $6$ magnitudes would move the water frost line to approximately $45$\,au, while during quiescence the frost line is
 closer to $4$\,au. This estimate has been recently confirmed by  \cite{2016arXiv160703757C}.
At 
 \EQ
 R=4, 9, 16 \text{ au},
 \EN
 the corresponding dimming rates in local orbits would be approximately
 \EQ
 \Lorb/L=8\%, 25\%, 50\%.
 \EN
Out to $10$\,au, those rates fall within the estimated bounds of \Eq{Lorblimits}, suggesting that preferential recondensation was
significant from Jupiter to Saturn, and possibly well beyond once the latent heat of water is taken into account.
 That suggests
 that as FU Orionis, or a similar object, fades significant preferential water ice recondensation occurs generating
 icy pebbles.

\section{Discussion and Conclusions}

The aerodynamics of dust grains, as measured through their Stokes number, plays into nearly every aspect of the formation
of and potentially also the growth of planetesimals \citep{2012A&A...544A..32L}. Preferential recondensation
naturally occurs in the aftermath of powerful accretion events such as FU Orionis type events,
providing a mechanism to create grains with thick enough icy mantles to be moderately decoupled from the gas ($St \gtrsim 0.01$); 
a potential observable. Further, different FU Orionis type objects, with differing cooling rates,
will have ice-mantled dust grains of differing sizes in their recently cooled regions.

Massive accretion events, FU Orionis outbursts occur early in the life-cycle of a protoplanetary disc with lots of gas left to play with, and are believed to be
a common phenomenon with most protostars undergoing several \citep{1996ARA&A..34..207H}.
While the radial extent of the accretion flow associated with the outburst is unclear, most of the energy is released at the disc's inner edge and it is
reasonable to assume a localized engine.
 We have shown that FU Orionis outbursts naturally
combine with preferential recondensation to provide a very rapid (orbital time scale) pathway to creating large ice-mantled dust grains. These
pebbles can be of the appropriate size to trigger
the Streaming Instability, leading to planetesimal formation very early in the protostar's life potentially at a large orbital separation \citep{2007Natur.448.1022J}. The
pebbles could also supply pebble accretion \citep{2015A&A...579A..43C},
allowing those early planetesimals to grow to become the cores of gas giants. Thus, evaporation and recondensation
could easily have played a major role in the formation of the gas giants in our own solar system; and could play
major roles in other forming planetary systems. This reinforces the concept of
intermittent thermal processing of solids in protoplanetary discs playing an important role in the process of planet formation \citep{2014Icar..237...84H}.

\section*{Acknowledgements}
The research leading to these results was funded by NASA OSS grant NNX14AJ56G.

\bibliography{Ices}

\end{document}